\documentclass[]{aastex6}

\fullcollaborationName{The Friends of AASTeX Collaboration}

\begin{document}

\title{VER J2227+608: A Hadronic PeVatron Pulsar Wind Nebula ?}

\author{Yuliang Xin\altaffilmark{1}, Houdun Zeng\altaffilmark{1}, Siming Liu\altaffilmark{1,2}, Yizhong Fan\altaffilmark{1,2} and Daming Wei\altaffilmark{1,2}}

\altaffiltext{1}{Key Laboratory of Dark Matter and Space Astronomy, Purple Mountain Observatory, Chinese Academy of Sciences,
Nanjing 210033, China;  ylxin@pmo.ac.cn, liusm@pmo.ac.cn}
\altaffiltext{2}{School of Astronomy and Space Science, University of Science and Technology of China, Hefei 230026, China}

\begin{abstract}

We report the detection of GeV $\gamma$-ray emission from the very-high-energy (VHE) $\gamma$-ray source VER J2227+608 associated with the ``tail'' region of SNR G106.3+2.7.
The GeV $\gamma$-ray emission is extended and spatially coincident with molecular clouds traced by CO emission.
The broadband GeV to TeV emission of VER J2227+608 can be well fitted by a single power-law function
with an index of 1.90$\pm$0.04, without obvious indication of spectral cutoff toward high energies.
The pure leptonic model for the $\gamma$-ray emission can be marginally ruled out by the X-ray and TeV data.
In the hadronic model, the low energy content of CRs and the hard $\gamma$-ray spectrum, in combination with the center-bright source structure, suggest that VER J2227+608 may be powered by the PWN instead of shocks of the SNR.
And the cutoff energy of the proton distribution needs to be higher than $\sim$ 400 TeV, 
which makes it an attractive PeVatron candidate.
Future observations by the upcoming Large High Altitude Air Shower Observatory (LHAASO) 
and the Cherenkov Telescope Array in the north (CTA-North) could distinguish these models and constrain the maximum energy of cosmic rays in supernova remnants.

\end{abstract}
\keywords{gamma rays: ISM - ISM: cosmic rays - individual: (G106.3+2.7 = VER J2227+608) - radiation mechanisms: non-thermal}

\section{Introduction} \label{sec:intro}

It is widely believed that supernova remnants (SNRs) are the dominant accelerators of Galactic cosmic rays (CRs).
The acceleration of electrons to hundreds of TeV energies has been supported by the detection of non-thermal synchrotron X-ray emission \citep{1995Natur.378..255K}. In the $\gamma$-ray band, {\em Fermi}-LAT and several ground-based Cherenkov telescopes, like HESS, VERITAS, MAGIC, etc, have detected tens of SNRs. Especially, the $\gamma$-ray emission of SNRs IC 443 and W44 observed by {\em Fermi}-LAT provides direct evidence for proton acceleration in SNRs \citep{2013Sci...339..807A}.
However, we still lack direct observational evidence for CRs in SNRs up to the spectral knee.
Pulsar wind nebulae (PWNe) driven by the central pulsars are also important particle accelerators and a large amount of PWNe have been detected in the $\gamma$-ray band \citep{2012A&A...548A..38A, 2018ApJ...867...55X}. 
High-energy particles can be accelerated at the termination shocks of PWNe.
In particular, GeV flares from the Crab nebula show that it can accelerate electrons to PeV energy \citep{2011A&A...527L...4B},
implying that PWNe could be important contributors to VHE Galactic CRs.

SNR G106.3+2.7 was first discovered by the northern Galactic plane survey of the Dominion Radio Astrophysical Observatory (DRAO) at 408 MHz\citep{1990A&AS...82..113J}. 
\citet{2000AJ....120.3218P} later found that it has a comet-shaped radio morphology with the spectral index $\alpha \approx 0.57 \pm 0.04$, where the flux density S$_{\nu} \propto \nu^{-\alpha}$. 
G106.3+2.7 consists of two distinct regions, a compact ``head'' region produced by the SNR shock interacting with the ambient dense material and an elongated ``tail'' region which corresponds to an outbreak to the interior of an neutral hydrogen (HI) bubble created by the stellar wind or supernova explosion \citep{2001ApJ...560..236K, 2006A&A...457.1081K}.
The ``head'' region also contains an off-center PWN in the north named ``Boomerang'' powered by the pulsar PSR J2229+6114, whose characteristic age and luminosity are 10 kyr and 2.2 $\times$ $10^{37}$erg s$^{-1}$, respectively \citep{2001ApJ...552L.125H}.
For the PWN, a spectral break at about 4.3 GHz is detected and has been attributed to the synchrotron cooling \citep{2006ApJ...638..225K}.
The ``tail'' region with a lower surface brightness also has a steeper spectrum with $\alpha \approx 0.70 \pm 0.07$ than the ``head'' region \citep[$\alpha \approx 0.49 \pm 0.05$;][]{2000AJ....120.3218P}.
\citet{2001ApJ...560..236K} detected HI and molecular materials associated with SNR G106.3+2.7 and derived a distance of 800 pc, which is much smaller than the value of 3 kpc estimated from X-ray absorption by \citet{2001ApJ...552L.125H}. Moreover, strong linearly polarized radio emission was detected in G106.3+2.7 \citep{2001ApJ...560..236K, 2011A&A...529A.159G}, which supports a close distance. At a distance of 800 pc, SNR G106.3+2.7 is 14 pc long and 6 pc wide.

In the $\gamma$-ray band, {\em Fermi}-LAT has detected GeV emission from PSR J2229+6114 \citep{2009ApJ...706.1331A}, which is associated with the previously unidentified EGRET source 3EG J2227+6122 \citep{1999ApJS..123...79H}.
Milagro has reported extended VHE $\gamma$-ray emission from G106.3+2.7 \citep{2007ApJ...664L..91A,2009ApJ...700L.127A}. 
However, it was not clear whether the emission origins from the SNR or pulsar complex.
In \citet{2009ApJ...703L...6A}, VERITAS collaboration detected significant TeV $\gamma$-ray emission from the elongated radio extension of SNR G106.3+2.7, $\sim 0.4 ^{\circ}$ away from the position of PSR J2229+6114 and named it VER J2227+608. 
The VHE energy spectrum can be well fitted by a power-law of the form $dN/dE = N_{0}(E/3 \rm TeV)^{-\Gamma}$, with an index of 
$\Gamma$ = 2.29 $\pm$ 0.33$_{\rm stat}$ $\pm$ 0.30$_{\rm sys}$ and the Milagro flux is coincide with the high-energy extrapolation of this spectrum \citep{2009ApJ...703L...6A}.
Meanwhile, the extended $\gamma$-ray emission is also coincident with molecular clouds traced by $^{12}$CO ($J$=1$-$0) emission \citep{1998ApJS..115..241H, 2001ApJ...560..236K}, favoring a hadronic origin of the $\gamma$-ray emission.
 
In this work, we carry out a detailed analysis of GeV emission near VER J2227+608 with ten years of {\em Fermi}-LAT Pass 8 data. In Section 2, the data analysis routines and results are presented. The origin of the $\gamma$-ray emission and the expected detection in the TeV band are discussed in Section 3, followed by conclusions in Section 4.

\section{Data Analysis} \label{sec:data}

\subsection{Data Reduction} \label{sec:data-reduction}

In the following analysis, we select the latest Pass 8 version of the {\em Fermi}-LAT data with ``Source'' event class (evclass=128 \& evtype=3), collected from August 4, 2008 (Mission Elapsed Time 239557418) to August 4, 2018 (Mission Elapsed Time 555033605). Only events with energies above 3 GeV are selected to avoid a too large point spread function (PSF) in the lower energy band. To suppress the contamination of the Earth Limb, the maximum zenith angle is adopted to be 90$^{\circ}$.
The region of interest (ROI) is a square area of $14^\circ \times 14^\circ$ centered at the position of VER J2227+608 \citep[R.A., decl.=$337.00^{\circ}$, $60.88^{\circ}$;][]{2009ApJ...703L...6A}.
We used the standard LAT analysis software {\it Fermitools}, available from the Fermi Science Support Center \footnote{http://fermi.gsfc.nasa.gov/ssc/data/analysis/software/}. 
Meanwhile, the binned likelihood analysis method with {\tt gtlike} and the instrument response function (IRF) of ``P8R3{\_}SOURCE{\_}V2'' are adopted to fit the data.
The diffuse backgrounds including the Galactic emission and the isotropic component,
are modeled according to {\tt gll\_iem\_v07.fits} and {\tt iso\_P8R3\_SOURCE\_V2\_v1.txt}\footnote {http://fermi.gsfc.nasa.gov/ssc/data/access/lat/BackgroundModels.html}.
All sources in the fourth Fermi catalog \citep[4FGL;][]{2019arXiv190210045T} within a radius of 20$^{\circ}$ from the ROI center and the two diffuse backgrounds are included in the source model.
During the fitting procedure, the spectral parameters and the normalizations of sources within distance smaller than $5^{\circ}$ to VER J2227+608 are set to be free, as well as the normalizations of the two diffuse backgrounds. 

\subsection{Results} \label{sec:data-results}

\begin{figure*}[!htb]
\centering
\includegraphics[width=3.0in]{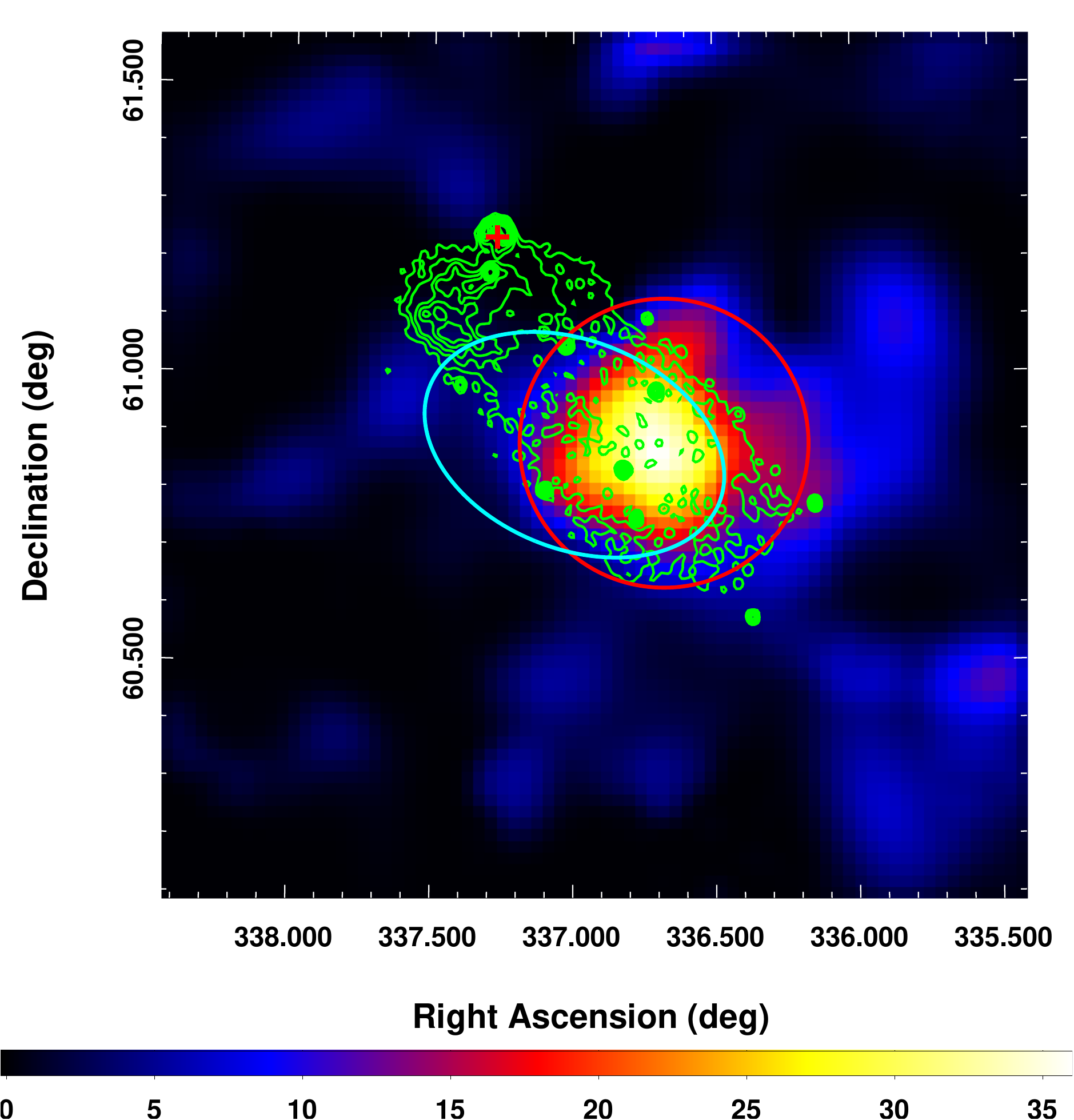}
\includegraphics[width=3.0in]{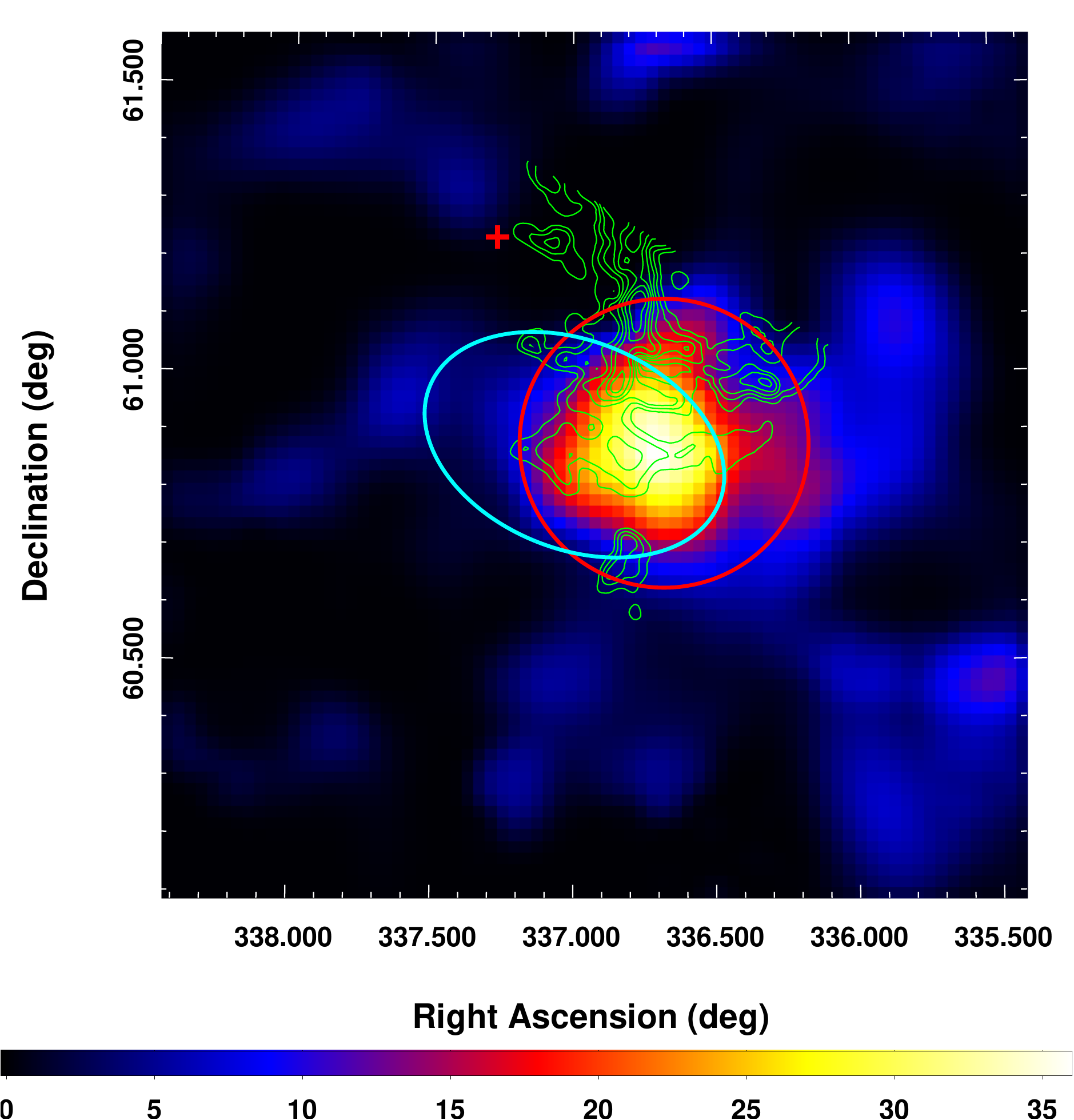}
\caption{The TS maps for a region of $2.0^\circ \times 2.0^\circ$ above 3 GeV with Gaussian smoothing of $\sigma$=0.04$^{\circ}$. In each panel, the position of 4FGL J2229.0+6114 associated with PSR J2229+6114 is marked by the red plus. The extension of SrcX is shown by the red circle and the cyan ellipse represents the extended TeV $\gamma$-ray emission of VER J2227+608\citep{2009ApJ...703L...6A}. The green contours in the left panel show the radio emission of SNR G106.3+2.7 at 1420 MHz by CGPS \citep{2001ApJ...560..236K, 2003AJ....125.3145T}. 
And in the right panel, the green contours display the $^{12}$CO ($J$=1$-$0) emission from FCRAO survey integrated between -5.59 and -7.23 km $\rm s^{-1}$ \citep{1998ApJS..115..241H, 2001ApJ...560..236K}.}
\label{fig:tsmap}
\end{figure*}

In the direction of SNR G106.3+2.7, there is only one 4FGL point source, 4FGL J2229.0+6114, which is the counterpart of PSR J2229+6114 \citep{2019arXiv190210045T}. We first created a $2.0^\circ \times 2.0^\circ$ Test Statistic (TS) map using {\em gttsmap} by subtracting the emission from the diffuse backgrounds and all 4FGL sources in the best-fit model, which is shown in the Figure \ref{fig:tsmap}. The TS map shows significant $\gamma$-ray emission toward the ``tail'' region of SNR G106.3+2.7, which is marked as SrcX afterwards. Then we added SrcX in the model file as a point source with power-law spectrum and redo the likelihood fitting. The best-fit position of SrcX optimized by the {\em gtfindsrc} command is R.A., decl.=$336.68^{\circ}$, $60.88^{\circ}$ with 1$\sigma$ error radius of $0.07^{\circ}$. 
The TS value of SrcX as a point source is 35.5, corresponding to a significance level of 5.1$\sigma$. 
The spectral index and integral photon flux are 1.60$\pm$0.17 and (9.18$\pm$2.68) $\times \rm 10^{-11}$ photon cm$^{-2}$ s$^{-1}$, respectively.

We carried out a spatial extension test for the $\gamma$-ray emission of SrcX using four uniform disks with different radii.
These uniform disks are centered at the best-fit position of SrcX and the radii of them are adopted to be $0.15^\circ$, $0.20^\circ$, $0.25^\circ$ and $0.30^\circ$. The TS values for the different spatial templates are 46.2, 49.8, 51.5 and 49.7, respectively.
The Akaike information criterion \citep[AIC;][]{1974ITAC...19..716A} is adopted to compare the different spatial models,
which shows $\Delta$AIC = AIC$_{\rm point}$ - AIC$_{\rm ext}$ = 14. Here AIC$_{\rm point}$ and AIC$_{\rm ext}$ represent the AIC values for point source and extended uniform disk with radius of $0.25^\circ$, respectively.
This result shows an obvious improvement for using the extended spatial template than a point-source hypothesis.
Meanwhile, the extended $\gamma$-ray emission of SrcX is consistent with the extended TeV $\gamma$-ray emission of VER J2227+608  as shown in Figure \ref{fig:tsmap} \citep{2009ApJ...703L...6A}, 
which supports SrcX to be the counterpart of VER J2227+608.
Hereafter, we adopted the uniform disk with radius of $0.25^\circ$ to describe the $\gamma$-ray emission of SrcX in the spectral analysis.
  
\begin{figure*}[!htb]
\centering
\includegraphics[width=3.5in]{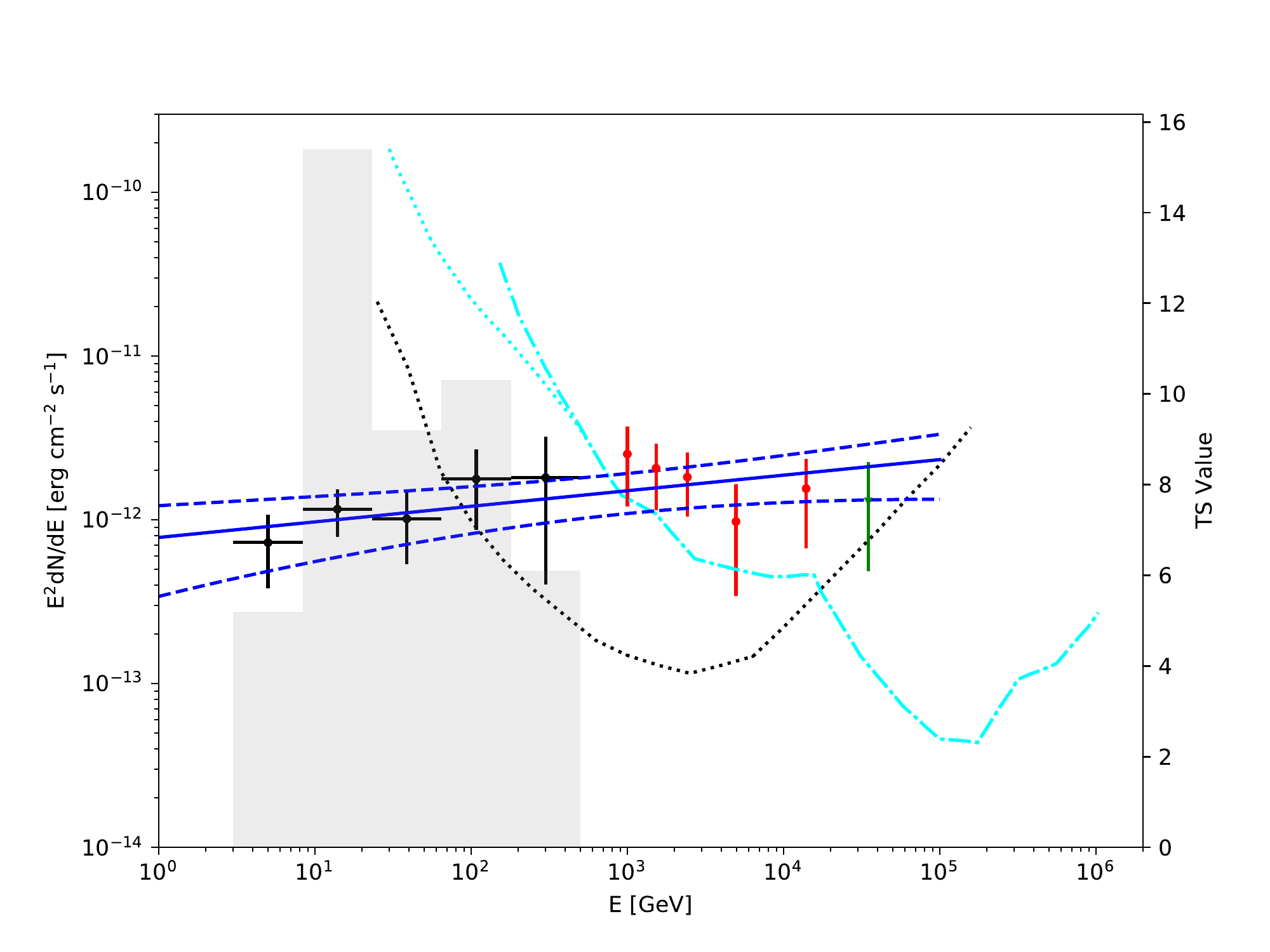}
\caption{The SED of SrcX marked by the black dots in the energy band of 3-500 GeV with the gray histogram shown as the TS value for each energy bin. The red and green dots show the TeV $\gamma$-ray data observed by VERITAS \citep{2009ApJ...703L...6A} and Milagro \citep{2007ApJ...664L..91A,2009ApJ...700L.127A}. A power-law spectrum with an index of 1.90$\pm$0.04 for the global $\gamma$-ray data is plotted by the blue solid line, and its 2$\sigma$ statistical error is marked as the blue dashed lines.
The cyan dotted and dot-dashed lines represent the differential sensitivities of LHAASO (1 year) with different sizes of photomultiplier tube \citep[PMT;][]{2019arXiv190502773B}. And the black dotted line shows the differential sensitivity of CTA-North \citep[50 hrs;][]{2019scta.book.....C}.}
\label{fig:sed}
\end{figure*}

With the extended spatial template for SrcX, the spectral index and integrated photon flux of SrcX in the energy band of 3-500GeV are fitted to be 1.81$\pm$0.16 and (2.04$\pm$0.48) $\times \rm 10^{-10}$ photon cm$^{-2}$ s$^{-1}$, respectively.
To obtain the spectral energy distribution (SED) of SrcX, the data are further divided into five equal logarithmic energy bins between 3 GeV and 500 GeV. And the likelihood analysis is repeated for each energy bin with only the spectral normalizations of
all sources within $5.0^\circ$ from SrcX and the two diffuse backgrounds left free.
For any bin with TS value of SrcX smaller than 5.0, a 95\% confidence level upper limit is calculated. 
The SED of SrcX is shown in Figure \ref{fig:sed}, together with the SED of VER J2227+608 \citep{2009ApJ...703L...6A}.
Interestingly, the GeV spectrum of SrcX is well consistent with the TeV spectrum of VER J2227+608 and the global spectral index is fitted to be 1.90$\pm$0.04 shown as the blue lines in Figure \ref{fig:sed}.

\section{Discussion} \label{sec:discussion}

The above data analysis shows that SrcX with the extended $\gamma$-ray emission is positionally coincident with VER J2227+608 and the GeV spectrum of SrcX is also in agreement with the TeV spectrum of VER J2227+608, which all support SrcX as the counterpart of VER J2227+608. 
In Figure \ref{fig:tsmap}, the radio emission of SNR G106.3+2.7 at 1420 MHz by the Canadian Galactic Plane Survey \citep[CGPS;][]{2001ApJ...560..236K, 2003AJ....125.3145T} shows that the $\gamma$-ray emission of VER J2227+608 is spatially consistent with the ``tail'' region of SNR G106.3+2.7. 
And in the right panel of Figure \ref{fig:tsmap}, the $^{12}$CO ($J$=1$-$0) emission integrated between -5.59 and -7.23 km $\rm s^{-1}$ from the Five College Radio Astronomy Observatory \citep[FCRAO;][]{1998ApJS..115..241H, 2001ApJ...560..236K} shows the molecular clouds around VER J2227+608, whose distance of 800 pc is derived from the center velocity of -6.4 km $\rm s^{-1}$ \citep{2001ApJ...560..236K}. 
This suggests that the $\gamma$-ray emission of VER J2227+608 may be associated with the molecular clouds.

For the $\gamma$-ray origin of VER J2227+608, several radiation mechanisms are considered here.
For the hadronic model, the $\gamma$-ray emission is from the decay of neutral pion mesons produced by inelastic \textit{pp} collisions. And for the leptonic model, the $\gamma$-ray emission can be produced by inverse Compton scattering (ICS) or a bremsstrahlung process of relativistic electrons. The radio and non-thermal X-ray emission is produced by high-energy electrons via the synchrotron process.

\citet{2000AJ....120.3218P} reported the radio flux densities and spectral indices of the ``tail'' and ``head'' regions of G106.3+2.7. And here we treat the ``tail'' region as the radio counterpart of VER J2227+608 considering the spatial correlation.
In the X-ray band, only PSR J2229+6114 and its PWN in the ``head'' region have been detected and the ``tail'' region has no significant X-ray emission \citep{2001ApJ...547..323H, 2001ApJ...552L.125H, 2002ASPC..271..199H}. Here, we adopted the intrinsic 2-10 keV flux of PSR J2229+6114, 1.3$\times$10$^{-12}$ erg cm$^{-2}$ s$^{-1}$ \citep{2001ApJ...552L.125H}, as an upper limit for VER J2227+608 in the X-ray band, which can be lower than that obtained via proper data analyses for the expended $\gamma$-ray source region.

Here we quantitatively discuss a pure leptonic model and a hadronic-leptonic hybrid one for the multi-wavelength spectrum of VER J2227+608. And the spectrum of electrons and/or protons is simply assumed to be a power-law spectrum with an exponential cutoff:
$dN/dE_i \propto E_i^{-\alpha_i} \exp(-E_i/E_{i,\rm cut})$, 
where $\alpha_i$ and $E_{i, \rm cut}$ are the spectral index and the cutoff energy, respectively, for $i = e$ or $p$.
The distance of VER J2227+608 is adopted to be 800 pc \citep{2001ApJ...560..236K}, and the radius of the radiation region is  $\sim$ 3.5 pc considering the best-fit uniform disk with radius of $0.25^\circ$ for the $\gamma$-ray emission in the data analysis above. Besides the cosmic microwave background (CMB), an infrared (IR) radiation field \citep[$T$ = 30 K, $u$ = 1.0 eV cm$^{-3}$;][]{2006ApJ...648L..29P} is also considered in the ICS process. The gas density is assumed to be $n_{\rm gas}$ = 1.0 cm$^{-3}$ for the bremsstrahlung and hadronic processes.

\begin{figure*}[!htb]
\centering
\includegraphics[width=7.0in,height=2.2in]{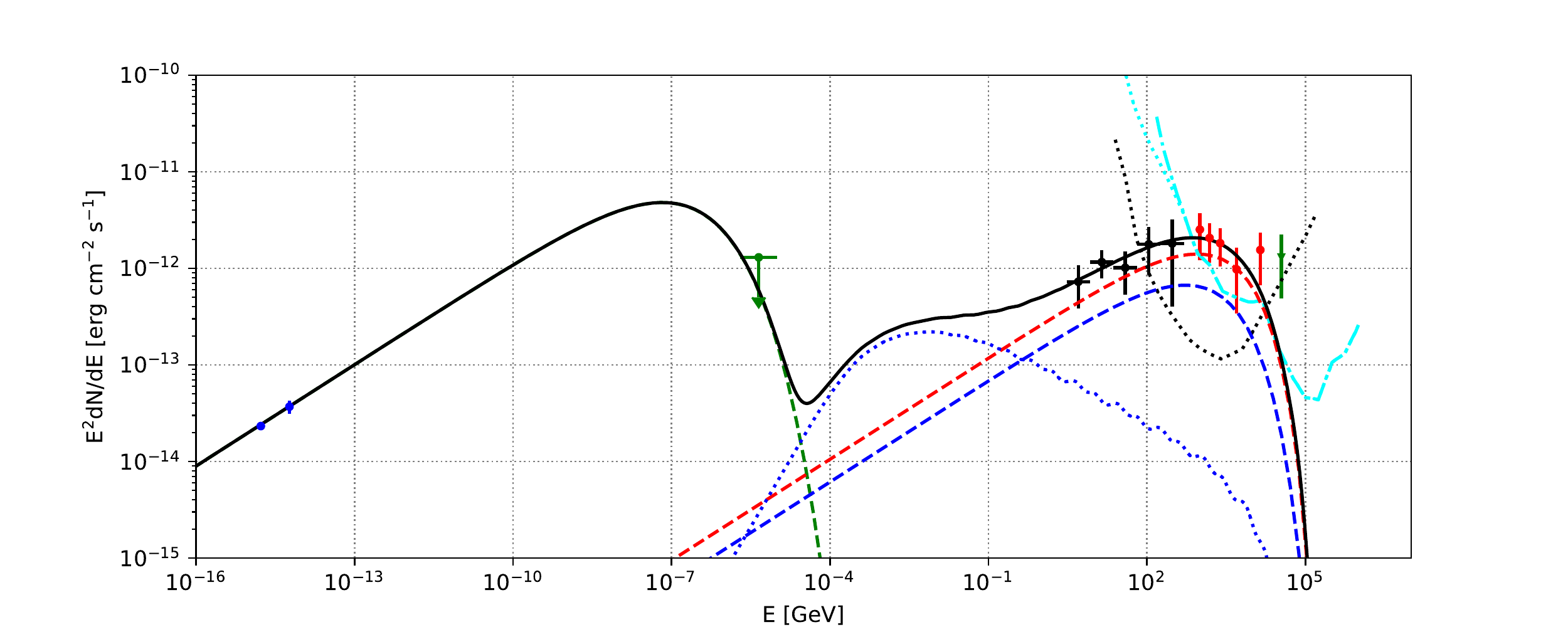}
\includegraphics[width=7.0in,height=2.2in]{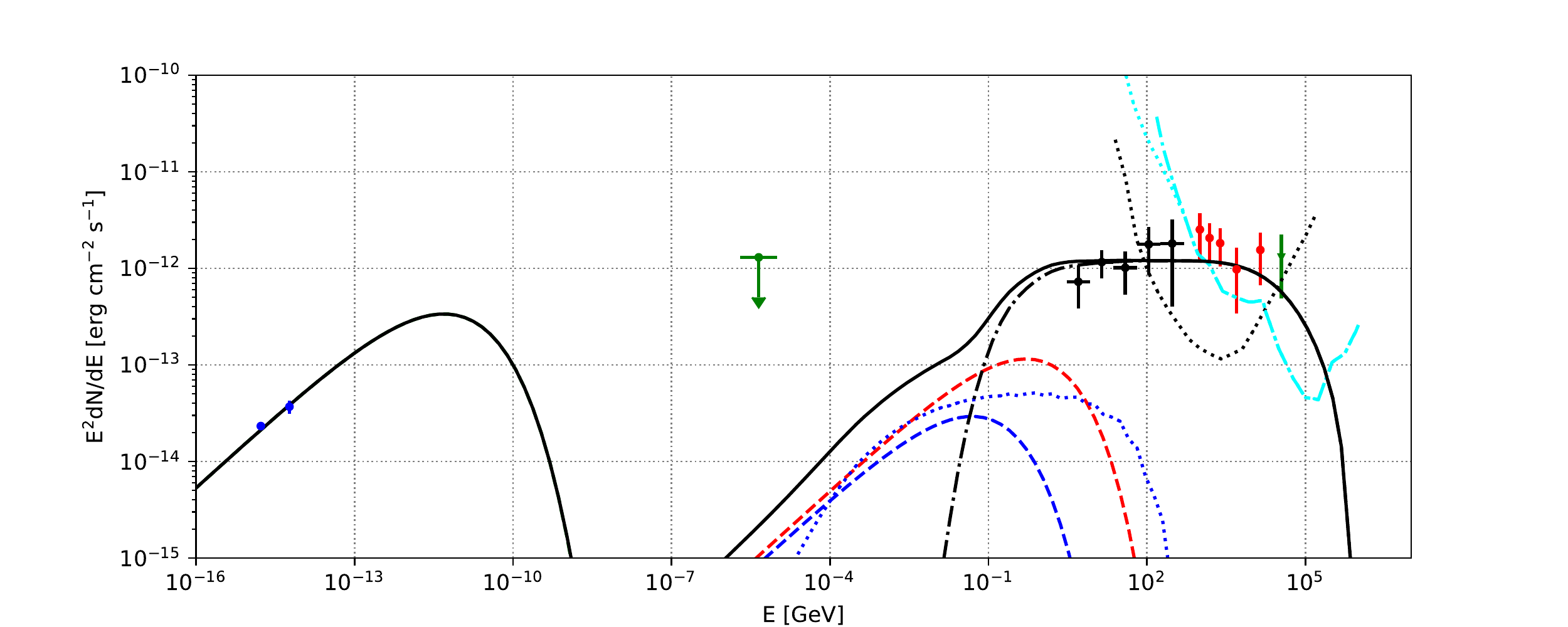}
\includegraphics[width=7.0in,height=2.2in]{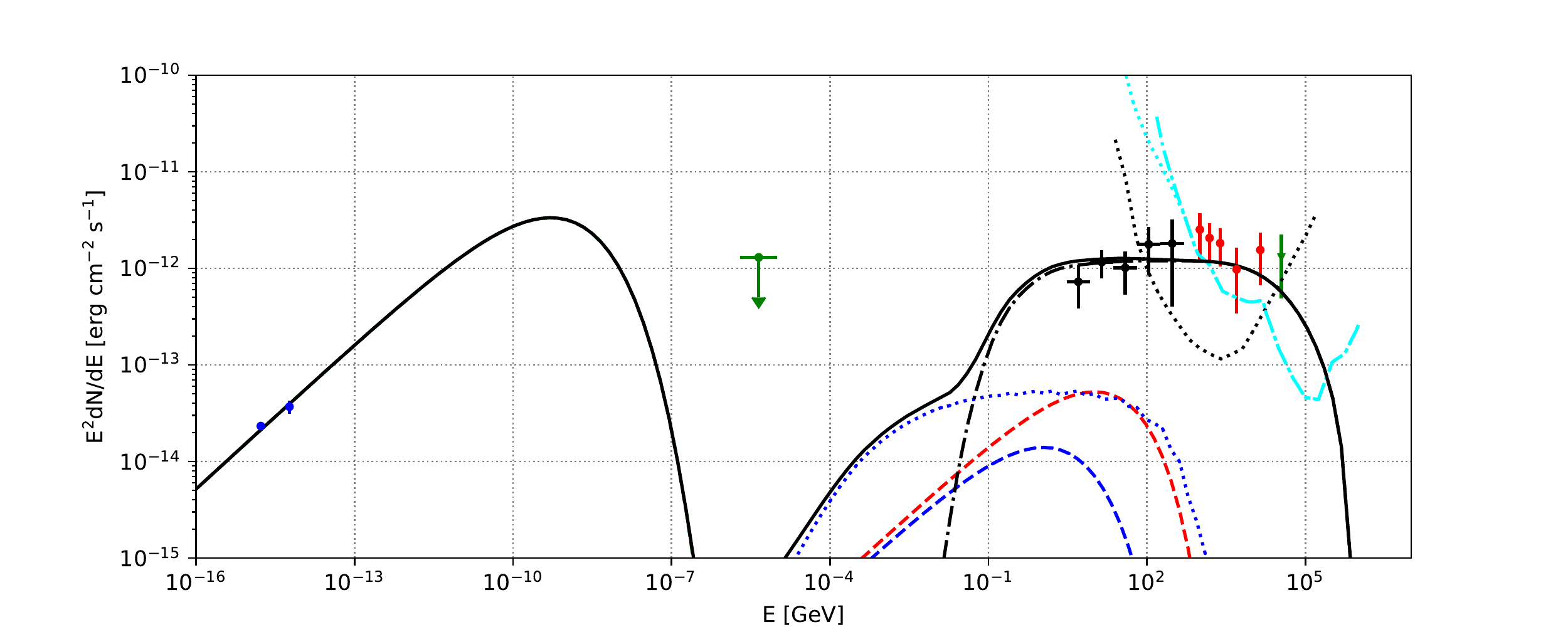}
\caption{Modeling of the multiwavelength SED of VER J2227+608. The top panel is for the leptonic model. The middle and bottom panels show the hybrid models with the gas density of $n_{\rm gas}$ = 1 cm$^{-3}$ and $n_{\rm gas}$ = 10 cm$^{-3}$, respectively.
In each panel, the green dashed line shows the synchrotron component. The bremsstrahlung emission is denoted by the blue dotted line and the black dot-dashed line presents the hadronic $\gamma$-ray emission. The blue and red dashed lines represent the ICS $\gamma$-ray emissions from scatterings off CMB and infrared photon fields, respectively. The sum of different radiation components is shown as the black solid line.}
\label{fig:model}
\end{figure*}

%differert models --- spectral expections --- LHAASO/CTA test
In the pure leptonic model, the spectral index of electrons is about 2.3. 
And the total energy of electrons above 1 GeV and the magnetic field strength are 4$\times$10$^{46}$ erg and 8 $\mu$G, respectively. Without considering the TeV data measured by Milagro \citep{2007ApJ...664L..91A,2009ApJ...700L.127A}, the cutoff energy of electrons need to be less than about 20 TeV to avoid exceeding the upper limit in the X-ray band.
With $B$ = 8 $\mu$G and $E_{\rm e,cut}$ = 20 TeV, the synchrotron cooling timescale of electrons is estimated to be about 9400 yr, which is close to the age of G106.3+2.7, $\sim 10^{4}$yr, inferred from the pulsar data \citep{2001ApJ...552L.125H}.
And such a scenario is likely consistent with the expectation about the evolution of high-energy particle distribution in SNRs
by \citet{2017ApJ...834..153Z,2019ApJ...874...50Z} and \citet{2019ApJ...876...24Z}, which suggests that the acceleration of high-energy electrons has stopped at present. However, this leptonic model can not reproduce the flux in the two highest energy bins and it requires that high-energy electrons accelerated by shocks in the early stage of the SNR evolution are still trapped inside the SNR.

%% Kep=0.01, n_gas=1cm-3
For the hadronic-leptonic hybrid model, the spectral indices of electrons and protons are adopted to be 2.0 to reduce the number of parameters. Meanwhile, the ratio of the normalization of electron distribution to that of protons, $K_{\rm ep}$, is fixed to be 0.01, which is in accord with the locally measured CR electron to proton flux ratio \citep{2012ApJ...761..133Y}.
With the gas density of $n_{\rm gas}$ = 1.0 cm$^{-3}$, the total energy of protons with the energy above 1 GeV is $\sim 6.0 \times 10^{48} (n/1.0\,\mathrm{cm}^{-3})^{-1}\ \mathrm{erg}$. To explain the flat $\gamma$-ray spectra, the cutoff energy of protons should be greater than 400 TeV. The cutoff energy of electrons should be less than $\sim$ 0.1 TeV to avoid excessive the flux in the GeV $\gamma$-ray band. And the magnetic field strength of $\sim 11 \mu$G is needed to explain the radio flux by synchrotron process.

%% Kep=0.01, n_gas=10cm-3
Considering the presence of molecular clouds in this region as shown in the right panel of Figure \ref{fig:tsmap}, we get an approximate density of about 10.0 cm$^{-3}$ based on the values of the column density, the CO brightness temperature, the velocity width and the approximate size of the molecular cloud \citep{2001ApJ...560..236K}. And we also present the hadronic-leptonic hybrid model with the gas density of $n_{\rm gas}$ = 10.0 cm$^{-3}$.
Here neutral gases and dense molecular clumps can serve as targets for hadronic processes.
With the cutoff energy of protons $\sim$ 400 TeV, the total energy content of protons above 1 GeV is reduced to be $\sim 6.0 \times 10^{47} (n/10.0\,\mathrm{cm}^{-3})^{-1}\ \mathrm{erg}$. Here the cutoff energy of electrons $E_{\rm e,cut}$ is poorly constrained and we fix $E_{\rm e,cut}$ by making the synchrotron radiation loss timescale equal to the age of SNR ($\sim$ 10$^{4}$ yr).
And the radio flux gives a magnetic field strength of $\sim 50 \mu$G and the corresponding $E_{\rm e,cut}$ is about 0.5 TeV.
Based on the observations of the HI and CO emission, \citet{2001ApJ...560..236K} suggested that the ``tail'' of G106.3+2.7 was created by a breakout of the SNR shock wave into the HI bubble with a lower density. 
However, the relatively lower energy content in CRs, the hard $\gamma$-ray spectrum, and its center bright morphology are distinct from typical SNRs \citep{2019ApJ...874...50Z}, implying that this source may be driven by the PWN instead of shocks of the SNR.
Nevertheless, if high-energy ions accelerated in the early stage of SNR evolution \citep{2017ApJ...844L...3Z} can be trapped within the SNR, the $\gamma$-ray emission may be attributed to the SNR.

If the tail is due to a breakout of the PWN into the HI bubble, we can get a value of the total ejected mass M$_{\rm ej}$ of the supernova using equation (2) in \citet{2004AdSpR..33..456C}, which describes the dynamical evolution of a PWN. 
And for PSR J2229+6114 with $\dot{E}$ = 2.2 $\times$ $10^{37}$erg s$^{-1}$ and the age of 10 kyr \citep{2001ApJ...552L.125H}, 
the value of M$_{\rm ej}$ is about 2.5 M$_\odot$, assuming a typical explosion energy of 10$^{51}$ erg and the rough radius 
of 14 pc (the long axis of G106.3+2.7). And this value of M$_{\rm ej}$ seems to be reasonable for typical SNRs \citet{2004AdSpR..33..456C}.
The $\gamma$-ray emission from PWN is usually attributed to the ICS process with electrons accelerated by its termination shock.
However, in dense environments, relativistic protons can also be produced and contribute to the $\gamma$-ray emission of PWN with
the hadronic process \citep{1990JPhG...16.1115C, 2008MNRAS.385.1105B}.
Although evidence for the presence of relativistic hadrons in PWNe has been elusive so far, 
the broadband spectra of several sources have been explained by the hadronic model, e.g. Vela X \citep{2009ApJ...699L.153Z},
PWN G54.1+0.3 \citep{2010MNRAS.408L..80L} and DA 495\citep{2019ApJ...878..126C}.
Moreover, \citet{2018MNRAS.478..926O} suggested that PWNe inside SNRs could further re-accelerate the relativistic protons up to the energy of 1 PeV and such PWNe may therefore be important PeVatrons.
The hard spectrum of protons and high proton cutoff energy of $>$400 TeV in the hybird model make VER J2227+608 a potential PeVatron. Other TeV sources with very hard spectra may have a similar origin \citep{2017ApJ...835...42G}.

Upcoming $\gamma$-ray experiments LHAASO \citep{2019arXiv190502773B, 2016ApJ...826...63L} and CTA-North \citep{2019scta.book.....C} can detect $\gamma$-ray emission from VER J2227+608.
As shown in Figure \ref{fig:sed}, the global spectral index of the $\gamma$-ray emission from VER J2227+608 is hard with spectral index of 1.90$\pm$0.04. And the differential sensitivities of LHAASO (1 year) with different sizes of photomultiplier tube \citep[PMT;][]{2019arXiv190502773B} and CTA-North \citep[50 hrs;][]{2019scta.book.....C} are also overplotted. Especially in the higher energy band (e.g. above 20 TeV), LHAASO has a great advantage than CTA-North. 
The different radiation models discussed above predict different spectra of VER J2227+608 for the energy above several tens of TeV, and the future accurate measurements of the TeV-PeV spectrum can distinguish these different radiation models and clarify whether it is a PeVatron.

\section{Conclusion} \label{sec:conclusion}

In this work, we analyze the $\gamma$-ray emission toward VER J2227+608, which is associated with the ``tail'' region of SNR G106.3+2.7, using ten years Pass 8 data of {\em Fermi}-LAT. The GeV $\gamma$-ray emission of VER J2227+608 is extended and the global GeV to TeV spectra can be described by a hard power-law form with an index of 1.90$\pm$0.04. 
Both the pure leptonic and the hadronic-leptonic hybrid models are considered for the multiwavelength data of VER J2227+608.
The pure leptonic model shows that the synchrotron cooling timescale of electrons is approximately equal to the age of SNR G106.3+2.7, implying quenching of VHE electron acceleration. However, the cutoff energy of electrons derived from the upper limit in the X-ray band seems to contradict with the TeV flux measured by Milagro.
The hybrid model seems to be favored considering the spatial coincident between the $\gamma$-ray emission and the likely dense molecular clouds traced by CO emission.
The cutoff energy of protons is constrained to be higher than $\sim$ 400 TeV.
Such high-energy protons makes VER J2227+608 a promising PeVatron candidate. 
The low energy content of CRs in this source, its hard $\gamma$-ray spectrum and center bright morphology 
suggest that it may be powered by the PWN instead of shocks of the SNR.
Future observations by LHAASO and CTA-North in the TeV band, especially accurate measurements of the TeV-PeV spectrum by LHAASO, would distinguish the different $\gamma$-ray origins of VER J2227+608 and test its PeVatron nature.

\acknowledgments

We would like to thank the anonymous referee for very helpful comments, which help to improve the paper significantly. 
This work is partially supported by the National Key R\&D Program of China 2018YFA0404203 and 2018YFA0404204, NSFC grants U1738122, U1931204 and 11761131007, the Natural Science Foundation for Young Scholars of Jiangsu Province, China (No. BK20191109),
and by the International Partnership Program of Chinese Academy of Sciences, grant No.114332KYSB20170008.



\end{document}